\documentclass[aps,pre,twocolumn,groupedaddress,showpacs]{revtex4}
\usepackage{graphicx}
\def\be{\begin{equation}}
\def\ee{\end{equation}}
\begin{document}
\title{Stability of off-axis motion for intense particle 
beams in periodically focusing channels}
\author{J.S. Moraes$^{a,b}$\footnote{e-mail: jsmoraes@if.ufrgs.br}, 
R. Pakter$^a$\footnote{e-mail: pakter@if.ufrgs.br}, and 
F.B. Rizzato$^a$\footnote{e-mail: rizzato@if.ufrgs.br}}
\affiliation{$^a$Instituto de F\'{\i}sica, Universidade Federal do Rio Grande do
Sul, Caixa Postal 15051, 91501-970 Porto Alegre, Rio Grande do Sul,
Brasil.}
\affiliation{$^b$Centro Universit\'ario La Salle,
Av. Victor Barreto, 2288, 92010-000, Canoas, RS, Brasil}
%

\begin{abstract}
A general equation for the centroid
motion of free, continuous, intense beams propagating off-axis
in solenoidal periodic focusing fields is derived.
The centroid equation
is found to be independent of the specific beam distribution and
may exhibit unstable solutions. A new Vlasov equilibrium 
for off-axis beam propagation is also obtained. Properties of the
equilibrium and the relevance of centroid motion to
beam confinement are discussed.

\end{abstract}
%
\pacs{41.85.Ja,05.45.-a}
\maketitle
%
A fundamental understanding of the kinetic equilibrium
and stability properties of high-current, low-emittance
beams in periodically focusing systems is crucial for
the development of a wide range of advanced
particle accelerator and coherent radiation source
applications. For a long time, the Kapchinskij-Vladimirskij
(KV) distribution \cite{kv59} was the only Vlasov equilibrium
distribution known for the propagation of periodically 
focused intense particle beams. Equilibrium and stability 
analysis based on the KV beam have been critical to the
development and understanding of the physics of 
intense beams 
\cite{hof83,struck84,chen94a,chen94b,gluck95,dav01,pak01,pak02,lund04}. 
More recently, it has been shown
that the KV distribution can be generalized to allow 
for rigid beam rotation with respect to the
Larmor frame in periodic solenoidal focusing
fields \cite{chen97}. Studies indicate that rotation
may have an important role in particle beam stability 
\cite{chen99}.

In the derivation of these Vlasov equilibria it is 
always assumed that the beam is perfectly aligned with 
the symmetry axis of the 
focusing field \cite{kv59,chen94b,chen97}. Actually,
this simplifying assumption is generally used in the 
analysis of intense beams \cite{dav01} 
because the axis is an equilibrium for the
beam centroid, and the equilibrium is stable if smooth-beam 
approximations are employed where the periodic 
fluctuations of the focusing field are
averaged out \cite{mark00}. In some cases, however,
we may expect the onset of parametric resonances involving the
centroid motion and the focusing field oscillations, which would
destabilize the centroid motion and heavily affect
the overall beam dynamics. In such conditions the
averaging procedure is no longer valid and a detailed description
of the centroid dynamics becomes mandatory.

In this paper, we derive from a kinetic Vlasov-Maxwell
description a general equation for the centroid
motion of free, continuous, intense beams propagating off-axis
in solenoidal periodic focusing fields. It is
shown that the centroid obeys a Mathieu type equation. 
The equation
is independent of the specific beam distribution and
becomes unstable whenever the oscillatory frequency 
of the centroid, which is related to the rms focusing
field strength per lattice, is commensurable with the 
focusing field periodicity itself. In the particular case
of a uniform beam density around the beam centroid, we
show that there exists a self-consistent Vlasov equilibrium 
distribution for the beam dynamics. The beam envelope
that determines de outer radius of the equilibrium beam around the
centroid is shown to obey the familiar envelope 
equation \cite{lapos71,sach71,chen94b,dav01}, 
being independent of the centroid motion. 
An example of the Vlasov equilibrium is discussed in detail to show
the possibility of finding beam solutions for which 
the extensively studied envelope equation 
\cite{struck84,chen94a,pak01,pak02,jor03,lund04} is stable,
whereas the centroid motion is unstable,
revealing the importance of the centroid motion to
overall
beam confinement properties.

We consider a free, continuous charged-particle beam propagating with average axial
velocity $\beta_bc{\bf \hat e}_z$ through a periodic solenoidal focusing 
magnetic field described by
\be
{\bf B}({\bf r},s)=B_z(s){\bf \hat e}_z-{r\over 2}B_z'(s){\bf \hat e}_r,
\label{b}
\ee
where ${\bf r}=x{\bf \hat e}_x+y{\bf \hat e}_y$,
$r=(x^2+y^2)^{1/2}$ is the radial distance from
the field symmetry axis, $s=z=\beta_bct$ is the
axial coordinate, $B_z(s+S)=B_z(s)$ is the
magnetic field on the axis, the prime denotes derivative
with respect to $s$, $c$ is the speed of light in {\it vacuo},
and $S$ is the periodicity length of the magnetic focusing field.
Since we are dealing with solenoidal focusing, it is convenient
to work in the Larmor frame of reference \cite{dav01}, which
rotates with respect to the laboratory frame with angular
velocity $\Omega_L(s)=qB_z(s)/2\gamma_bmc$, where
$q$, $m$ and $\gamma_b=(1-\beta _b^2)^{-1/2}$ are,
respectively, the charge, mass and relativistic factor of
the beam particles. The Larmor frame is used throughout the paper,
such that ${\bf \hat e}_x$ and ${\bf \hat e}_y$ are assumed to
be versors along the rotating axes.
In the paraxial approximation, the beam distribution 
function $f_b({\bf r},{\bf v},s)$ evolves according 
to the Vlasov-Maxwell system \cite{dav01}
\begin{eqnarray}
{\partial f_b \over \partial s} + {\bf v} \cdot \nabla f_b +(-\kappa_z {\bf r}
-\nabla
\psi) \cdot \nabla_{\bf v} f_b = 0, \label{ap11} \\
\nabla^2 \psi = - (2 \pi K/N_b) \> n_b({\bf r},s), \label{ap12}\\
n_b=\int f_b d{\bf v}, \label{ap13}
\end{eqnarray}
where $n_b({\bf r},s)$ is the beam density profile,
$\kappa_z(s) =q^2B^2_z(s)/4\gamma_b^2\beta_b^2m^2c^4$
is the focusing field parameter, 
$K=2q^2N_b/\gamma_b^3\beta_b^2mc^2$ is
the beam perveance, $N_b=\int f_bd{\bf r}d{\bf v}=$const. 
is the number of
particles per unit axial length, and ${\bf v}\equiv {\bf r}'$ \cite{note}.
In Eqs. (\ref{ap11})-(\ref{ap13}),
$\psi$ is a normalized potential that incorporates both
self-electric and self-magnetic fields, ${\bf E}^s$ and
${\bf B}^s$. It is related
to the self-scalar and self-vector potentials by
$\phi^s=\beta_b^{-1}A_z^s=\gamma_b^3m\beta_b^2c^2\psi({\bf r},s)/q$,
where ${\bf A}^s({\bf r},s)=A_z^s({\bf r},s){\bf \hat e}_z$,
${\bf E}^s({\bf r},s)=-\nabla \phi^s({\bf r},s)$, and
${\bf B}^s({\bf r},s)=\nabla\times{\bf A}^s({\bf r},s)$.

Our first task here is to determine the evolution of the
beam centroid located at
\be
\bar {\bf r}(s)\equiv N_b^{-1} \int {\bf r} 
f_b({\bf r},{\bf v},s)
d{\bf r} d{\bf v}.
\ee
In order to do that one multiplies Eq.~(\ref{ap11}) 
by ${\bf r}$ and integrates over phase-space
to get
\begin{equation}
\bar{\bf r}' = \bar{\bf v},
\label{ap2}
\end{equation}
where $\bar{\bf v} \equiv N_b^{-1} \int {\bf v} f
d{\bf r} d{\bf v}$. If one now multiplies 
Eq.~(\ref{ap11}) by ${\bf v}$ and integrates over 
phase-space, one obtains
\begin{equation}
\bar{\bf v}' = -\kappa_z \bar{\bf r} - \overline{\nabla \psi},
\label{ap3}
\end{equation}
where $\overline{\nabla \psi} \equiv N_b^{-1} \int \nabla \psi f d{\bf r} d{\bf
v}$ is obtained by integration by parts of the $\nabla_{\bf v}$ - term in
velocity space. Using Eqs. (\ref{ap12}) and (\ref{ap13}) we can 
rewrite $\overline{\nabla \psi}$
as
\begin{equation}
\overline{\nabla \psi} =
(2 \pi K)^{-1} \int \nabla \psi \nabla^2 \psi d{\bf r}.
\label{ap4}
\end{equation}
Then we note that the integrand of Eq. (\ref{ap4}) can be
cast into the more suitable form
\begin{equation}
\nabla \psi \nabla^2 \psi =
\nabla \cdot [\nabla \psi \nabla \psi - {\bf I} (\nabla \psi)^2/2]
\label{ap5}
\end{equation}
where
the unit dyadic ${\bf I}$ reads ${\bf I} \equiv {\bf \hat e}_x{\bf \hat e}_x
+{\bf \hat e}_y{\bf \hat e}_y$. Now, employing Gauss theorem we
obtain
\be
\overline{\nabla \psi} =
(2 \pi K)^{-1} \oint {\bf \hat e}_n\cdot
\left[\nabla \psi \nabla \psi - {\bf I} (\nabla \psi)^2/2\right]
dA=0,
\label{oint}
\ee
because
$\nabla \psi \rightarrow 0$ as $r \rightarrow \infty$ for beams in free
space. In Eq. (\ref{oint}), $dA$ and ${\bf \hat e}_n$ are, respectively,
the boundary differential element and the unit vector normal to the boundary of 
integration located at $r \rightarrow \infty$.
In fact, the result $\overline{\nabla \psi}=0$ is expected based
on the action-reaction law; since $-\nabla \psi$ corresponds
to the self-force exerted on the beam particles by themselves, its
average throughout the beam distribution has to vanish due to
the pairwise structure of the interparticle electromagnetic interaction.
Using Eqs. (\ref{ap3}) and (\ref{oint}) in Eq. (\ref{ap2}), we finally
obtain the centroid equation of motion
\begin{equation}
\bar{\bf r}'' + \kappa_z (s)\bar{\bf  r}=0.
\label{centroid}
\end{equation}
Let us stress that we have not made any assumption on the
particular form of the beam distribution function so far.
Thus, the centroid equation above is always valid as
long as the beam evolves according to the Vlasov-Maxwell
system, Eqs. (\ref{ap11})-(\ref{ap13}).
In the laboratory frame, combined to the oscillatory 
motion described by Eq. (\ref{centroid}) the centroid
also rotates with angular velocity $\Omega_L(s)$ around
the center $r=0$.
Taking into account that
$\kappa_z(s)$ is periodic, Eq. (\ref{centroid}) is of
the Mathieu type which is known to present
unstable solutions related to parametric resonances
in the $\bar{\bf r}$ motion. If we conveniently write the
average of $\kappa_z(s)$ over one lattice period as
$(1/S) \int_0^S \kappa_z(s) ds \equiv \sigma_0^2/S^2$,
where $\sigma_0$ is a dimensionless parameter proportional to the 
rms focusing field measuring the vacuum phase advance in
the small field, smooth-beam approximation, the
instabilities in the centroid motion are expected when one 
approaches $\sigma_0 \sim n\pi$; this condition corresponds
to parametric resonances between the oscillation periodicity
of $\bar{\bf r}$ in the average (rms) focusing field and the
periodicity of the focusing field itself. Depending
on the exact profile of $\kappa_z(s)$ the size of the unstable
regions surrounding $\sigma_0 \sim n\pi$ vary significantly.
If the aim is beam confinement, these regions are to be avoided.

It is worth mentioning that although Eq. (\ref{centroid}) is strictly 
valid for free beams only, it
is expected to provide a good description of the centroid motion
in bounded systems if the beam is nearly symmetric and is not 
excessively displaced from a
pipe center located at $r=0$. 
The reason is because in this case $\nabla \psi = \pm {\bf \hat e}_n |\nabla
\psi|$ at the pipe walls, where ${\bf \hat e}_n$ is now the unit 
vector normal to the
wall, and the surface integral in Eq. (\ref{oint}), performed
along the boundary, still vanishes 
since $|\nabla\psi|$ is approximately constant there. Note
also that the presence of a pipe would generally not suppress the 
centroid instabilities discussed in connection with Eq. (\ref{centroid}); 
in fact, it would even enhance it because
the image charges induced are of opposite sign, 
attracting the beam to the wall.

Our next task is to show that we can construct a Vlasov
equilibrium for off-axis beam transport. In particular, 
we assume
a beam with a uniform radial density distributed around a center
located at 
${\bf r_o}(s)=x_o(s){\bf \hat e}_x+y_o(s){\bf \hat e}_y$, 
i.e.,
\be
n_b({\bf r},s)=\cases{N_b/\pi r_b^2(s),& $r_\delta<r_b(s),$\cr
0,& $r_\delta>r_b(s),$}
\label{nb}
\ee
where $r_b(s)$ is the equilibrium beam envelope and
${\bf r_\delta}\equiv{\bf r}-{\bf r_o}$. A schematic of the
beam distribution of Eq.~(\ref{nb}) and corresponding
vectors is shown in Fig. 1. For such beam we can easily recognize
${\bf r_o}(s)$ as being the centroid coordinate. According to what was
shown previously, its evolution must then obey
\be
{\bf r_o}''+\kappa_z (s){\bf  r_o}=0.
\label{ro}
\ee
Using the prescribed $n_b({\bf r},s)$ in Eq.~(\ref{ap12}) 
we find for the normalized self-potential
\be
\psi({\bf r},s)=-Kr_\delta^2/2r_b^2(s)
\ee
in the beam interior ($r_\delta<r_b$). 
Therefore, a single
particle of the beam located at ${\bf r}(s)$
subjected to the external focusing field force
$-\kappa_z(s){\bf r}$ and the self-field force 
$-\nabla\psi({\bf r},s)$ will evolve according to
\be
{\bf r}''+\kappa_z\,{\bf r}-(K/ r_b^2)\, {\bf r_\delta}=0.
\label{r}
\ee
If we now subtract Eq. (\ref{ro}) from Eq. (\ref{r}) we obtain
\be
{\bf r_\delta}''+\kappa_z\,{\bf r_\delta}-(K/ r_b^2)\,{\bf r_\delta}=0,
\label{9}
\ee
which describes the motion of the beam particle with respect to the center
${\bf r_o}$. Equation (\ref{9}) can be solved with known 
techniques of physics of beams \cite{chen94b,dav01}. 
Considering the motion along the $x$-axis, we write 
$x_\delta=A_{x_\delta} w(s) \sin[\int^s\zeta(s)ds+\zeta_{x_\delta0}]$
with $A_{x_\delta}$ and $\zeta_{x_\delta0}$ constants. Substituting
this expression into Eq. (\ref{9}) we obtain
\be
w'' + \kappa(s) w = w^{-3},
\label{10}
\ee
$\zeta(s) = w^{-2}(s)$, where $\kappa(s) \equiv \kappa_z(s) - K/r_b^2(s)$,
and the constant of motion $A_{x_\delta}$ can be expressed in the form
\begin{equation}
A_{x_\delta}^2=(x_\delta/ w)^2 + (w x_\delta '- w'x_{\delta})^2.
\label{11}
\end{equation}
Performing an equivalent calculation for the motion along
the $y$-axis, one shows that $A_{y_\delta}$ given by
\be
A_{y_\delta}^2=(y_\delta/ w)^2 + (w y_\delta '- w'y_{\delta})^2.
\ee
is also a constant of motion. From Eq. (\ref{9})
one sees that all the forces are
central with respect to the centroid ${\bf r_o}$. Thus, one readily 
demonstrates
that the canonical angular momentum $P_{\Theta\delta}$ given by
\be
P_{\Theta\delta}=x_\delta y_\delta '- y_\delta x_\delta '
\ee
is a constant of motion as well.
Because $A_{x_\delta}^2$, $A_{y_\delta}^2$, and $P_{\Theta\delta}$
are exact single-particle constants of motion, a
possible choice of Vlasov equilibrium distribution function
is
\begin{eqnarray}
f_b^{EQ}({\bf r},{\bf v},s)={N_b\over\pi^2\epsilon_T} 
\delta\Big[A_{x_\delta}^2+A_{y_\delta}^2-\nonumber\\
2\omega_bP_{\Theta\delta}
-\left(1-\omega_b^2\right)\epsilon_T\Big],
\label{feq}
\end{eqnarray}
where $df_b^{EQ}/ds=0$, $\epsilon_T=$const is an
effective emittance, and the rotation parameter $\omega_b=$const is
in the range $-1<\omega_b<1$ for radially confined beams. Using
$f_b^{EQ}$ in Eq. (\ref{ap13}), it is
readily shown that the uniform density profile centered at
${\bf r_o}$ of Eq. (\ref{nb}) is consistently obtained, provided
$r_b(s)=\epsilon_T^{1/2}w(s)$. Hence, $r_b(s)$ obeys the
familiar envelope equation
\begin{equation}
r_b'' +\kappa_z(s) r_b - {K \over r_b} - {\epsilon_T^2 \over r_b^3}=0.
\label{rb}
\end{equation}
Performing the appropriate averages over the equilibrium 
distribution, Eq. (\ref{feq}), we can show that
the beam rigidly rotates around its centroid 
${\bf r_o}(s)$ with angular velocity 
$\Omega_{b\delta}(s)=\omega_b\epsilon_T\beta_bc/r_b^2(s)$.
Also, the rms emittance calculated
with respect to the centroid position is given by
\begin{equation}
\epsilon_\chi\equiv 4\left[\left\langle \chi ^2\right\rangle 
\left\langle \chi '^2\right\rangle-
\left\langle \chi \ \chi'\right\rangle^2\right]^{1/2}=\epsilon_T,
\label{12.5}
\end{equation}
where $\chi=x_\delta,y_\delta$, and the brackets indicate
averages over the beam distribution.
One thus sees that a Vlasov equilibrium distribution can be formed 
for which the beam envelope obeys Eq.~(\ref{rb}) with constant emittance even when
the centroid moves off-axis, ${\bf r}_o \neq 0$, following the dynamics dictated by
Eq. (\ref{ro}). We refer to this equilibrium as a {\it periodically focused
off-axis Vlasov equilibrium}. Let us call attention to the interesting 
fact that the centroid motion and the envelope dynamics are
uncoupled in this case. 
In other words, centroid dynamics does not affect the known stability
results for the envelope dynamics \cite{pak01,struck84,jor03,lund04,chen94a} 
and is not affected 
by the latter as well. One should keep in mind that for good beam 
confinement both centroid and
envelope have to be stable.

\begin{figure}
\vskip .2cm
\includegraphics[scale=.3]{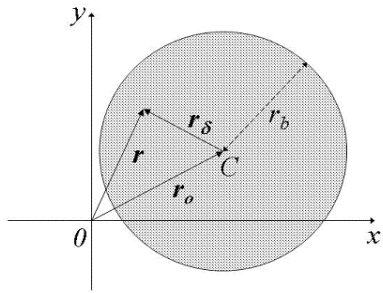}
\caption{Schematic of the
beam distribution of Eq.~(\ref{nb}) and corresponding
vectors. $C$ corresponds to the centroid position.}
\end{figure}

\begin{figure}[h]
\vskip -.3cm
\includegraphics[scale=.4]{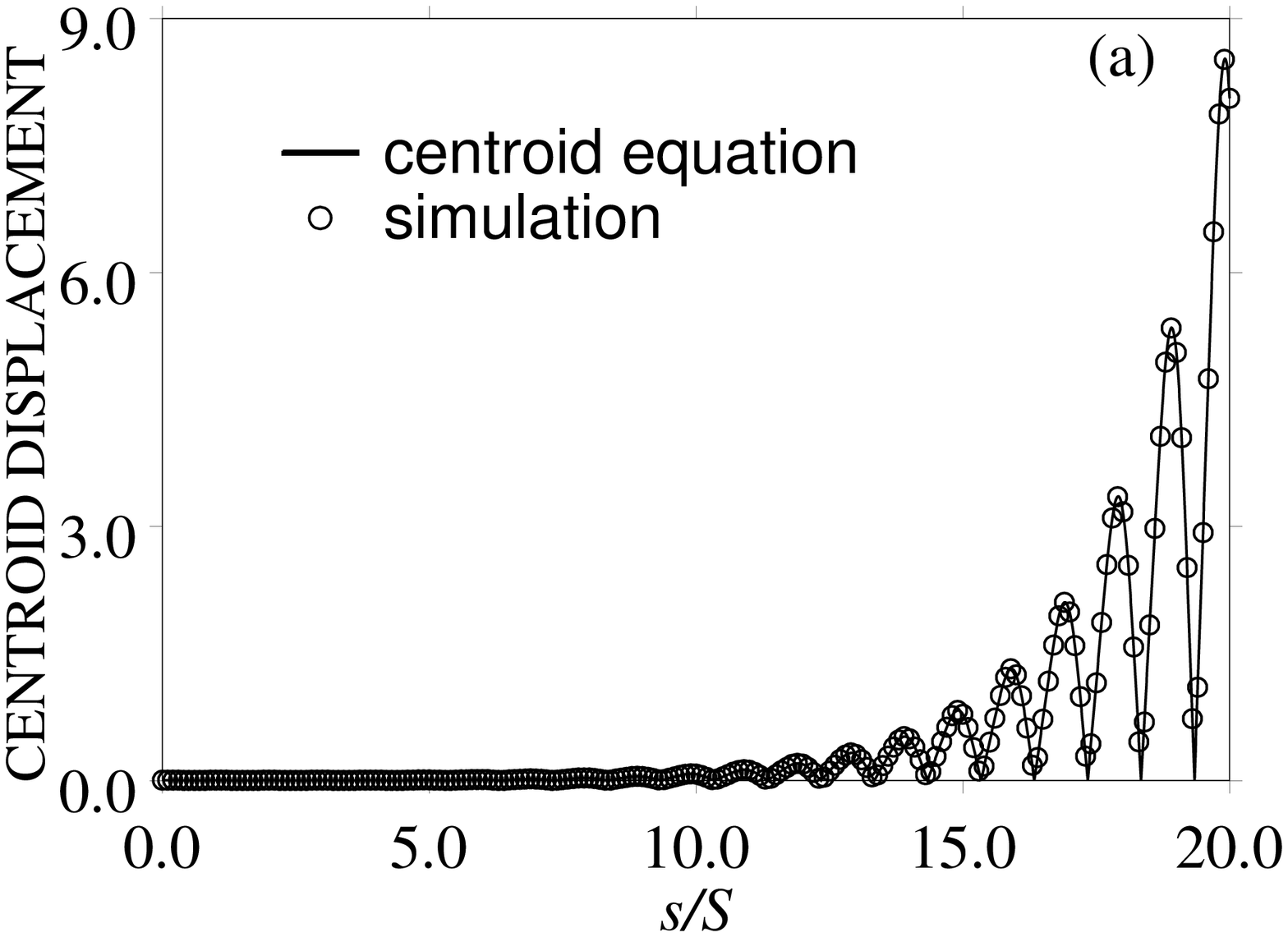}
\vskip -.4cm
\hskip .2cm
\includegraphics*[scale=.4,trim=.5cm 0 0 0]{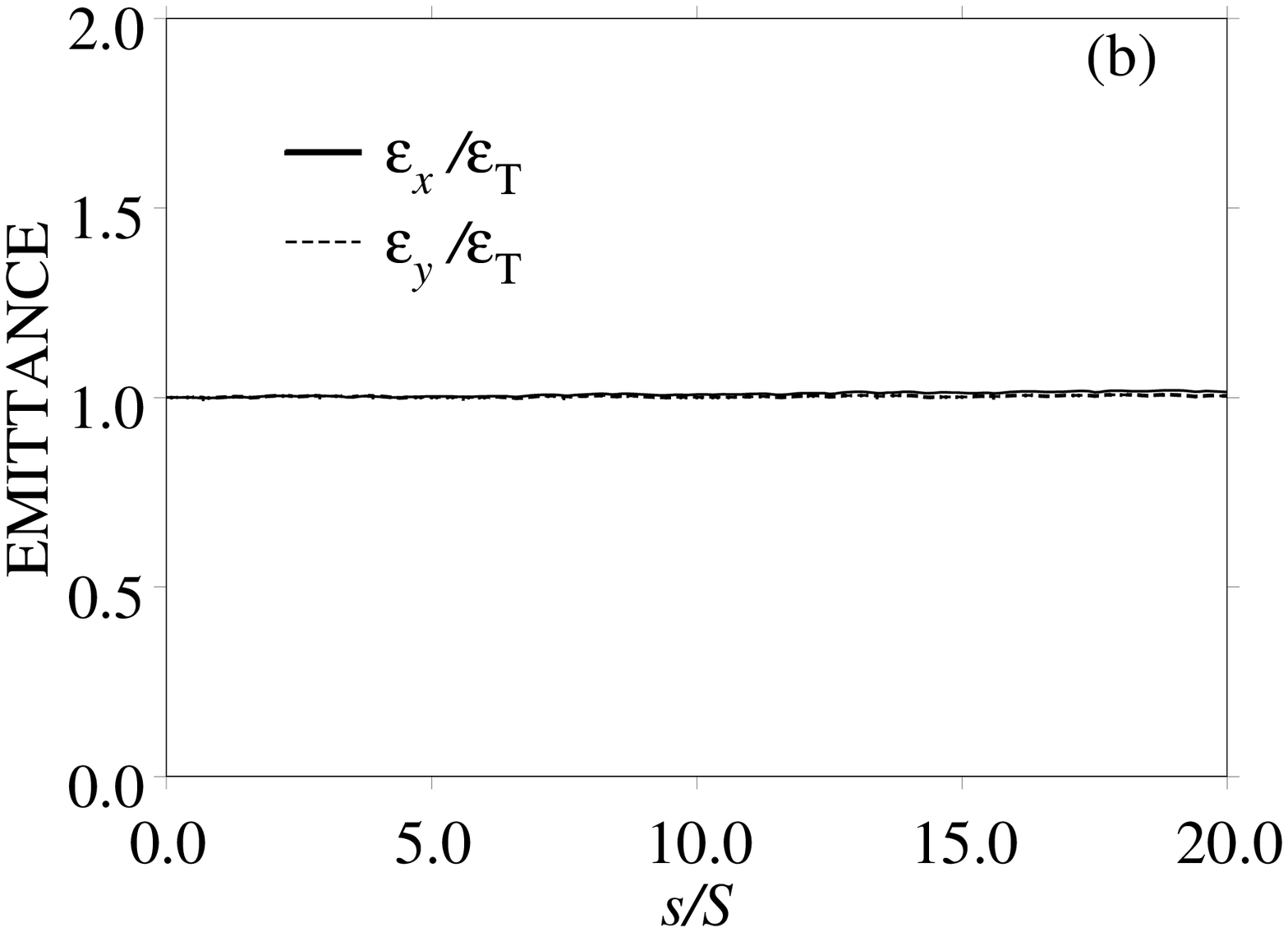}
\vskip -.4cm
\hskip .2cm
\includegraphics*[scale=.4,trim=.5cm 0 0 0]{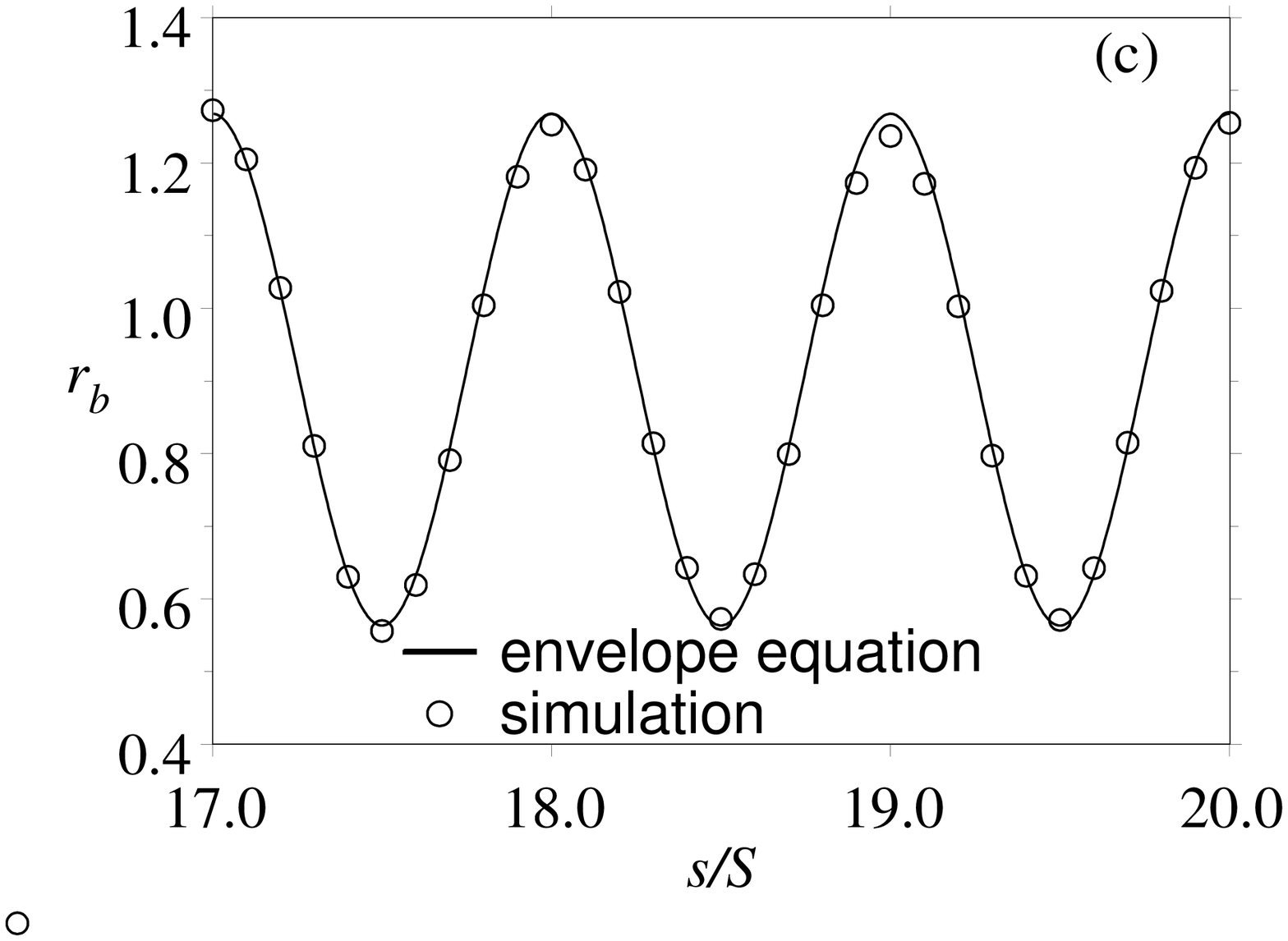}
\vskip -.4cm
\caption{Multiparticle self-consistent simulation results. 
(a) The centroid motion; (b) rms emittance; and (c) the envelope dynamics.
Centroid displacement and envelope are normalized to $(S\epsilon_T)^{1/2}$.}
\end{figure}

We now illustrate our results with an example of periodically focused
off-axis Vlasov equilibrium. 
We consider a particular set of parameters for which
the envelope equation (\ref{rb}) is known to be stable,
whereas the centroid motion of Eq. (\ref{ro}) was found
to be unstable. 
We investigate beam transport with the aid of
self-consistent numerical simulations, where a large number
$N_b=8000$ of macroparticles interact via pairwise electromagnetic 
interactions \cite{pak02}.
In the simulation we used $SK/\epsilon_T=5.0$ and 
$S^2\kappa_z(s)=\sigma_0^2[1+\cos(2\pi s/ S)]$,
with $\sigma_0=155^o$, over 20 lattice periods. 
The macroparticles were launched at $s=0$ according to
the equilibrium distribution, Eq.~(\ref{feq}), with $\omega_b=0$, 
${\bf r_o}=0= {\bf r_o}'$, and $r_b$ corresponding to the matched solution
with $r_b(s+S)=r_b(s)$ of the envelope equation (\ref{rb}).
The finite number of macroparticles in the initial 
condition acts as a seed for any possible instability to develop.
Simulation results are presented
in Fig. 2. The evolution of the centroid displacement $r_o\equiv|{\bf r_o}|$
calculated from the macroparticles positions ${\bf r}$ as
${\bf r_o}=<{\bf r}>$, where the brackets
indicate average over macroparticles, is shown in
Fig. 2(a) (circles). It reveals that the centroid motion develops the typical
exponential growth of unstable dynamics that agrees with the fact that
the set of parameters considered leads to an unstable solution
of Eq. (\ref{ro}). The solid line corresponds to the
solution obtained from Eq. (\ref{ro}).
Despite the centroid instability, the beam equilibrium 
distribution is preserved as verified in Fig. 2(b) that shows that 
rms emittance is well conserved as the beam evolves. RMS emittance is
calculated according to Eq. (\ref{12.5}), considering averages over
macroparticles. Finally, Fig. 2(c) compares the envelope obtained 
from the envelope equation (\ref{rb}) with that obtained from the
simulation, $r_b = [2 <({\bf r}-{\bf r_o})^2>]^{1/2}$, for the
last 3 periods of the focusing channel. 
The perfect agreement proves once more the preservation of 
the equilibrium distribution. Moreover, we see that in spite of
the unstable centroid the envelope is stable,
as predicted.

To conclude, based on kinetic grounds we have derived
a general equation for the centroid
motion of free, continuous, intense beams propagating off-axis
in solenoidal periodic focusing fields. It was
shown that the centroid equation
is independent of the specific beam distribution and
may exhibit unstable solutions. In the particular case
of a uniform beam density around 
the beam centroid, we have
shown the existence of a periodically focused
off-axis Vlasov equilibrium 
distribution describing a beam that rigidly rotates
with a prescribed angular velocity around a moving centroid. 
The beam envelope
around the centroid was shown to obey the familiar envelope 
equation, 
being independent of the centroid motion. 
An example of periodically focused
off-axis Vlasov equilibrium was discussed in detail to show
the possibility of finding beam solutions for which 
the envelope equation is stable,
whereas the centroid motion is unstable,
revealing the importance of centroid motion to
the overall beam confinement properties.

\smallskip
We acknowledge partial support from CNPq, Brazil.


\end{document}